\documentstyle[aps,eqsecnum,preprint,tighten]{revtex}

\begin{document}

\draft


%
\preprint{\vbox{Submitted to Physical Review D \hfill TRI-PP-95-13\\}}
\title{Baryon QCD sum rules in an external isovector-scalar field\\
and baryon isospin mass splittings}
\author{Xuemin Jin}
\address{TRIUMF, 4004 Wesbrook Mall, Vancouver,
British Columbia, Canada V6T 2A3\\}
\date{\today}
\maketitle
\begin{abstract}
Within the QCD sum-rule approach in an external field,
we calculate the baryon matrix element of isovector-scalar
current, $H_{\rm B}=\langle B|\overline{u}u-
\overline{d}d|B\rangle/2M_{\rm B}$, for octet baryons,
which appears in the response of the correlator
of baryon interpolating fields to a constant isovector-scalar
external field. The sum rules are obtained for a general
baryon interpolating field with an appropriate form for
the phenomenological ansatz of the spectral density.
 The key phenomenological input is
the response of the quark condensates to the external field.
To first order in the quark mass difference $\delta m=m_d-m_u$,
the non-electromagnetic part of the baryon isospin mass
splitting is given by the product of $\delta m$ and
$H_{\rm B}$. Therefore, QCD sum-rule calculation
of $H_{\rm B}$ leads to an estimate of the octet baryon
isospin mass splittings.
The resulting values are comparable to the
experimental values; however, the sum-rule predictions
for $H_{\rm B}$ are sensitive to
the values of the response of the quark condensates
to the external source, which are not well determined.

\end{abstract}


\section{Introduction}
\label{intro}

To understand the observed properties of hadrons from
the underlying theory of the strong interaction, quantum
chromodynamics (QCD),  is a challenging task
since the QCD remains intractable at low energies.  Among the
attempts made in dealing with the strong interactions
at low energy scales is the QCD sum-rule approach\cite{svz1},
which has proven to be a useful tool of extracting
qualitative and quantitative information about hadronic
 properties\cite{svz1,reinders1}.

One of the extension of the sum-rule methods made by
Ioffe and Smilga\cite{ioffe1}
for external field problems enables one to calculate the
baryon matrix elements of various bilinear quark operators. These
include the matrix element of electromagnetic current to determine the
magnetic moments\cite{ioffe1,balitsky1,chiu1}, the matrix element of
the axial vector current to find the renormalization of baryon axial
coupling constant\cite{belyaev1,chiu2}, the matrix element of the
quark part of the energy momentum tensor, which gives the momentum
fraction carried by the up and down quarks in deep inelastic
scattering\cite{kolesinchenko1,belyaev2}, and the matrix element of
isoscalar-scalar current for evaluating the nucleon
sigma term\cite{jin1}.

In this paper, we evaluate the baryon matrix elements of
isovector-scalar current, $H_{\rm B}=
\langle B|\overline{u}u-\overline{d}d|
B\rangle/2M_{\rm B}$, within the external field QCD sum-rule
approach. In Ref.~\cite{jin2}, the
proton matrix element $\langle p|\overline{u}u-\overline{d}d|
p\rangle/2M_p$ has been calculated in the external field approach.
However, one piece of the phenomenological representation has been
omitted in the calculation. This has been pointed out
recently by Ioffe\cite{ioffe2}. In the present paper, we shall
derive the appropriate phenomenological representation, which
includes the piece neglected in Ref.~\cite{jin2}. We use this
complete phenomenological representation and a general
baryon interpolating field to calculate the matrix element
$H_{\rm B}$ for octet baryons.

External field sum rules for baryons are based on the study of
the correlation function of the baryon interpolating
field in the presence of an external field.
The appearance of the external field leads to specific new features in
QCD sum rules  which distinguish them from those in the absence of the
external field. At the hadron level, the spectral parameters usually
used in the parametrization of the spectral density, baryon masses,
pole residues, and continuum thresholds, all respond to
the external field. Consequently the phenomenological representation
for the
{\it response} of the correlation function contains a double pole at the
baryon mass whose residue contains the matrix element of interest.
This corresponds to the response of the pole position.
The response of the pole residues gives rise to
single pole terms, which contain the
information about the transition between the ground state
baryon and excited states. The single pole contributions are not
exponentially damped after Borel transformation relative to the double
pole term and should be retained in a consistent analysis of the sum
rules. In addition, there are terms corresponding to the
response of the continuum thresholds, which should also be
included in the calculation. At the quark level,
the external field contributes in
two different ways--by directly coupling to the quark fields in the
baryon current and by polarizing the QCD vacuum.
By equating these two different representations for
the response of the baryon correlator, one obtains the
external field sum rules, which relate the baryon matrix
elements of various current to QCD Lagrangian parameters,
vacuum  condensates, and the response of condensates to
the external source.

The observed baryon isospin mass splitting has its origin in the
electromagnetic interactions between quarks and in the different masses of
the up and down current quarks. The contributions of the latter to the
first order in the quark mass difference $\delta m=m_d-m_u$ is given by
the product of $\delta m$ and the baryon matrix element of the
isovector-scalar current. Therefore, the QCD sum-rule calculation
of the baryon matrix elements of isovector-scalar current
for octet baryons naturally
leads to an estimate of the octet baryon isospin mass splittings.
(The $\Sigma^--\Sigma^0$
and $\Sigma^+-\Sigma^0$ splittings will not be considered here as there is
mixing of the $\Sigma^0$ with the $\Lambda$ via isospin-violating
interactions.)

The rest of this paper is organized as follows. In Sec.~\ref{sumrule},
we establish the baryon QCD sum rules in an external isovector-scalar
field. In Sec.~\ref{anay}, we then analyze the sum rules and present
the results. In Sec.~\ref{isospin}, we estimate the isospin mass
splittings of the octet baryons using the baryon matrix elements
of isovector-scalar current calculated from QCD sum rules. Further
discussion of our results are given in Sec.~\ref{discussion}.


\section{Baryon QCD sum rules in an external isovector-scalar field}
\label{sumrule}

In this section, we establish the baryon QCD sum rules in
the presence of an external isovector-scalar field.
In previous
works\cite{ioffe1,balitsky1,chiu1,belyaev1,%
chiu2,kolesinchenko1,belyaev2,jin1,jin2,ioffe2},
the phenomenological representation
for the correlator is usually obtained by analyzing
a double dispersion relation. Here we present an
alternative approach to derive the phenomenological
representation.
The operator product expansion (OPE) results
can be easily obtained following the procedures outlined in
Ref.~\cite{jin2}.
 We work to leading order in perturbation theory and to first
order in the strange quark mass. Contributions proportional to
the up and down current quark masses and the gluon condensate
are neglected as they
give numerically small contributions. We include condensates
up to dimension eight.

Consider the correlator of the baryon interpolating field in
the presence of a {\it constant} external isovector-scalar field
$S_{\rm V}$
\begin{equation}
\Pi(S_{\rm V},q)\equiv i\int d^4{x}e^{iq\cdot x}
\langle 0|{{\rm T}[\eta_{\rm B}(x)\overline{\eta}_{\rm B}(0)]}\rangle_{S_
{\mbox{\tiny{\rm V}}}}\ ,
\label{corr}
\end{equation}
where $\eta_{\rm B}$ is the interpolating field for the baryon under
consideration.
We consider baryon interpolating fields (currents) that contain no
derivatives and couple to spin-$1\over 2$ states only.  There are two
linearly independent fields with these features, corresponding to a
scalar or pseudoscalar diquark coupled to a quark.
In this paper, we take a linear combination of these two fields,
\begin{eqnarray}
\eta_p(x)&=&2\epsilon_{abc}
\left\{[u_a^T(x) C d_b(x)]\gamma_5 u_c(x)+
t [u_a^T(x) C \gamma_5 d_b(x)]u_c(x)\right\}\ ,
\label{eta-p}
\\*[7.2pt]
\eta_{\Sigma^{+}}(x)&=&2\epsilon_{abc}
\left\{[u_a^T(x) C s_b(x)]\gamma_5 u_c(x)+
t [u_a^T(x) C \gamma_5 s_b(x)]u_c(x)\right\}\ ,
\label{eta-s}
\\*[7.2pt]
\eta_{\Xi^0}(x)&=&2\epsilon_{abc}
\left\{[s_a^T(x) C u_b(x)]\gamma_5 s_c(x)+
t [s_a^T(x) C \gamma_5 u_b(x)]s_c(x)\right\}\ ,
\label{eta-x}
\end{eqnarray}
where $u(x)$, $d(x)$ and $s(x)$ stand for the up, down and
strange quark fields, $a,b$ and $c$ are the color indices,
$C=-C^T$ is the charge conjugation matrix, and  $t$ is an
arbitrary real parameter.
The interpolating fields for neutron, $\Sigma^{-}$ and
$\Xi^{-}$ can be obtained by changing $u(d)$ into $d(u)$.
The interpolating fields with
$t=-1$, advocated by Ioffe \cite{ioffe3,ioffe4},
have been used exclusively
in previous papers on external field sum rules
\cite{ioffe1,balitsky1,chiu1,belyaev1,%
chiu2,kolesinchenko1,belyaev2,jin1,jin2,ioffe2}.
In principle, the sum-rule predictions
are independent of the choice of $t$; in practice, however, the OPE is
truncated and the phenomenological description is represented roughly.
The goals in choosing the interpolating field for QCD sum-rule
applications are to maximize the coupling of the interpolating field to
the state of interest relative to other (continuum) states, while
minimizing the contributions of higher-order terms in the OPE.  These
goals cannot be simultaneously realized.
The optimal choice of the baryon interpolating field seems to be around
Ioffe's choice.  We refer the reader to Refs.~\cite{ioffe4,leinweber1}
for more discussion about the choice of baryon interpolating fields.
We shall consider the interval $-1.15\leq t\leq
-0.85$ here. For $t>-0.85$ the continuum contributions become large
while for $t<-1.15$  the contributions from higher-order terms in the
OPE become important relative to the leading-order terms.

The subscript $S_{\rm V}$ in Eq.~(\ref{corr}) indicates the
presence of the external field. Thus,
the correlator should be calculated with an additional term
\begin{equation}
\Delta{\cal L} \equiv - S_{\rm V} [\overline u (x) u(x)
-\overline d(x) d(x)]\; ,
\label{lag}
\end{equation}
added to the usual QCD Lagrangian, and $-\Delta{\cal L}$
added to ${\cal H}_{\rm QCD}$.
Since $S_{\rm V}$ is a scalar constant,
Lorentz covariance and parity allow one to decompose
$\Pi(S_{\rm V},q)$ into two distinct structures\cite{jin2}
\begin{equation}
\Pi(S_{\rm V},q)\equiv \Pi^1(S_{\rm V},q^2)+
\Pi^q(S_{\rm V},q^2)\rlap{/}{q}\ .
\end{equation}
To obtain QCD sum rules, one needs to construct a phenomenological
representation for $\Pi(S_{\rm V},q)$ and evaluate
$\Pi(S_{\rm V},q)$ using the OPE.

\subsection{The dispersion relation and phenomenological spectral ansatz}

To determine the correlator at the hadron level we
use the dispersion relation
\begin{equation}
\Pi^i(S_{\rm V},q^2)=\int_0^\infty{\rho^i(S_{\rm V},s)
\over s-q^2}ds\
\label{des-rel}
\end{equation}
for each invariant function $\{i=1,q\}$, where
$\rho^i(S_{\rm V},s)={1\over\pi}{\rm Im}\Pi^i(S_{\rm V},s)$
is the spectral density. Here we have omitted polynomial
subtractions which will be eliminated by a subsequent Borel
transformation.
We have also omitted infinitesimal as we are only concerned
with large and space like $q^2$ in QCD sum rules.

In practical applications of QCD sum-rule approach, one usually
parametrizes the spectral density by a simple pole representing the
lowest energy baryon state of interest plus a continuum which
is approximated by a perturbative evaluation of the correlator
starting at an effective threshold\cite{svz1,reinders1,ioffe3}.
When $S_{\rm V}$
is present, we add $-\Delta{\cal L}$ to ${\cal H}_{\rm QCD}$,
which is equivalent to increase $m_u$ and $m_d$ by $S_{\rm V}$
and $-S_{\rm V}$, respectively.
Consequently at the hadron level, the baryon spectrum will
be shifted. Since we are concerned here with the linear response
to the external source, $S_{\rm V}$ can be taken to be arbitrarily
small (see below). Thus, there
is no rearrangement of the spectrum, and we can use a pole
plus continuum ansatz for the baryon spectral density
\begin{equation}
\rho^i(S_{\rm V},s)=\lambda_{\rm B}^{*^2}\phi^i \delta (s-M_{\rm B}^{*^2})
+\widetilde{\rho}^i(S_{\rm V},s)\theta (s-s_0^{*^i})\ ,
\label{ans}
\end{equation}
where $\phi^{*^i}=\{M^*_{\rm B}, 1\}$ for $\{i=1,q\}$, and
$\widetilde{\rho}^i(S_{\rm V},s)$ is to be evaluated in
perturbation theory.
Here $\lambda^*_{\rm B}$ is  defined by
$\langle 0|\eta_{\rm B}|{\rm B}\rangle_{S_{\rm V}}
=\lambda_{\rm B}^* v^*_{\rm B}$
with $v^*_{\rm B}$ the Dirac spinor normalized to
$\overline{v}^*_{\rm B} v^*_{\rm B}=2M^*_{\rm B}$, $M^*_{\rm B}$
is the mass of the lowest baryon state and $s_0^{*^i}$ is the
continuum threshold in the presence of the external field.

Let us now expand both sides of Eq.~(\ref{des-rel}) for small $S_{\rm V}$
\begin{equation}
\Pi^i_0(q^2)+S_{\rm V}\Pi^i_1(q^2)+\cdots
=\int_0^\infty {\rho^i_0(s)\over s-q^2} ds
+S_{\rm V}\int_0^\infty {\rho^i_1(s)\over s-q^2} ds+\cdots\ .
\label{des-expand}
\end{equation}
Since $S_{\rm V}$ is arbitrary, one immediately concludes that
\begin{eqnarray}
\Pi^i_0(q^2)&=&\int_0^\infty{\rho^i_0(s)\over s-q^2} ds\ ,
\label{des-2}
\\*[7.2pt]
\Pi^i_1(q^2)&=&\int_0^\infty {\rho^i_1(s)\over s-q^2} ds\ .
\label{des-3}
\end{eqnarray}
Obviously, Eq.~(\ref{des-2}) leads to the baryon {\it mass} sum rules
in vacuum which have been extensively studied
\cite{ioffe3,belyaev3,reinders1,leinweber2}. Here we are interested in
Eq.~(\ref{des-3}), which corresponds to the linear response of the
correlator to the external source and contains the baryon
matrix element under consideration (see below).

Expanding the right-hand side of Eq.~(\ref{ans}), we find
\begin{eqnarray}
\rho^i_0(s)&=&\lambda^2_{\rm B}\phi^i_0\delta(s-M^2_{\rm B})+
\widetilde{\rho}^i_0(s)\theta(s-s_0^i)\ ,
\label{ans-0}
\\*[7.2pt]
\rho^i_1(s)&=&-2 H_{\rm B}\, M_{\rm B}\lambda^2_{\rm B} \phi_0^i
\delta^\prime (s-M^2_{\rm B})+\Delta\lambda_{\rm B}^2\,\phi_0^i
\delta (s-M^2_{\rm B})
\nonumber
\\*[7.2pt]
& &
+\Delta\phi^i\,  \lambda^2_{\rm B}\delta (s-M^2_{\rm B})
-\Delta s_0^i\,  \widetilde{\rho}^i_0(s)\delta(s-s_0^i)+
\widetilde{\rho}^i_1(s)\theta (s-s_0^i)\ ,
\label{ans-1}
\end{eqnarray}
where we have defined
\begin{eqnarray}
M^*_{\rm B}&=&M_{\rm B}+S_{\rm V} H_{\rm B}+\cdots\ ,
\\*[7.2pt]
\lambda^{*^2}_{\rm B}&=&\lambda^2_{\rm B}
+S_{\rm V} \Delta\lambda^2_{\rm B}+\cdots\ ,
\\*[7.2pt]
s_0^{*^i}&=&s_0^i+S_{\rm V} \Delta s_0^i+\cdots\ ,
\\*[7.2pt]
\phi^{*^i}&=&\phi^i_0+S_{\rm V}\Delta\phi^i+\cdots\ ,
\\*[7.2pt]
\widetilde{\rho}^{*^i}(s)&=&\widetilde{\rho}^i_0(s)+S_{\rm V}
\widetilde{\rho}^i_1(s)+\cdots\ ,
\end{eqnarray}
where the first terms are the vacuum spectral parameters
in the absence of the external field.
Note that $\Delta\phi^1=H_{\rm B}$ and $\Delta\phi^q=0$.
Treating $S_{\rm V}$ as a small parameter, one can
 use the Hellman-Feynman theorem\cite{hellman1,feynman1}
 to show that
\begin{equation}
H_{\rm B}={\langle B|\overline{u}u-\overline{d}d|B\rangle
\over 2M_{\rm B}}\ ,
\label{hf-s}
\end{equation}
where we have used covariant normalization $\langle k^\prime,
B|k, B\rangle=(2\pi)^2 k^0\delta^{(3)}(\vec k^\prime-\vec k)$.

One notices that $\rho^i_1(s)$ has specific new features
which distinguish it from $\rho^i_0(s)$. The first term
in Eq.~(\ref{ans-1}), which is {\it absent} in $\rho^i_0(s)$,
give rise to a double pole at the baryon mass whose residue
contains the matrix element of interest. The second and
third terms are single pole terms; the residue at the single
pole contains the information about the
transition between the ground state baryon and the excited states.
In terms of quantum mechanical perturbation, the double pole term
corresponds to the energy shift while the single pole
terms result from the response of baryon wave function to
the external field.  The fouth term is due to the response
of the continuum threshold to the external source and the
last term is the continuum contribution. As emphasized in
the previous works, the
single pole contributions are not exponentially damped after
the Borel transformation relative to the double
term and should be retained in a consistent analysis of the sum
rules.

The fourth term has been neglected in Ref.~\cite{jin2}.
The contribution
of this term is suppressed in comparison with the single
pole terms by a factor $e^{-(s_0^i-M^2_{\rm B})/M^2}$
[see Eqs.~(\ref{sum_p_1}--\ref{sum_x_q})].
If the response of the continuum threshold is small, one can neglect
the contribution of the fourth term. However, if
the response of the continuum threshold is strong, one needs
to include the fourth term
in the calculation. This point has been noticed recently by
Ioffe in Ref.~\cite{ioffe2}, where a double dispersion
relation is considered for the vertex function
\begin{equation}
\Pi_1(q)=\int d^4 x e^{iq\cdot x}\langle 0|T
\eta_{\rm B}(x)\left[\int d^4 z (\overline{u}(z)u(z)-\overline{d}(z)d(z))
\right]\overline{\eta}_{\rm B}(0)|0\rangle\
\label{vertex}
\end{equation}
in order to get the appropriate phenomenological representation.
[This vertex function can be obtained by  expanding the
right-hand side of Eq.~(\ref{corr}) directly.]
We note that our discussion and
Eq.~(\ref{ans-1}) are consistent with those given
in Ref.~\cite{ioffe2}.
Substituting Eq.~(\ref{ans-1}) into Eq.~(\ref{des-3}),
one obtains the appropriate phenomenological representation.


\subsection{QCD representation}

The QCD representation of the correlator is obtained by applying
the OPE to the
time-ordered product in the correlator. When the external field
is present, the up and down quark fields satisfy the modified
equations of motion:
\begin{eqnarray}
(i\rlap{\,/}D-m_u-S_{\rm V})u(x)&=& 0\ ,
\\*[7.2pt]
(i\rlap{\,/}D-m_d+S_{\rm V})d(x)&=& 0\ ,
\label{eq-mo}
\end{eqnarray}
where $\rlap{\,/}D=\gamma^\mu(\partial_\mu-ig_s{\cal A}_\mu)$ is
the covariant derivative. (The equation of motion for the
strange quark field does not change.)
In the framework of the OPE, the external field contributes to
the correlator in two ways: It couples directly to
the quark fields in the baryon interpolating fields and it also
polarizes the QCD vacuum. Since the external
 field in the present problem is a Lorentz scalar,
non-scalar correlators cannot be induced in the QCD vacuum. However the
external field does modify the condensates already
present in the QCD vacuum. To first order in $S_{\rm V}$,
the chiral quark condensates
can be written as follows
\begin{eqnarray}
\langle\overline{u}u\rangle_{S_{\mbox{\tiny{\rm
 V}}}}&=&\langle\overline{u}u\rangle_{\mbox{\tiny{\rm 0}}}-\chi
 S_{\mbox{\tiny{\rm V}}}\langle\overline{u}u\rangle_{\mbox
{\tiny{\rm 0}}}\ ,
\label{uc}
\\*[7.2pt]
\langle\overline{d}d\rangle_{S_{\mbox{\tiny{\rm
 V}}}}&=&\langle\overline{d}d\rangle_{\mbox{\tiny{\rm 0}}}+\chi
S_{\mbox{\tiny{\rm V}}}\langle\overline{d}d\rangle_{\mbox{\tiny
{\rm 0}}}\ ,
\label{dc}
\\*[7.2pt]
\langle\overline{s}s\rangle_{S_{\mbox{\tiny{\rm
 V}}}}&=&\langle\overline{s}s\rangle_{\mbox{\tiny{\rm 0}}}-\chi_{\rm s}
S_{\mbox{\tiny{\rm V}}}\langle\overline{s}s\rangle_{\mbox{\tiny
{\rm 0}}}\ ,
\label{sc}
\end{eqnarray}
where $\langle \hat{O}\rangle_{\mbox{\tiny{\rm 0}}}\equiv \langle
0|\hat{O}|0\rangle$.
%
%
The mixed quark-gluon condensates change in a similar way
\begin{eqnarray}
\langle g_s\overline{u}\sigma\cdot {\cal G} u
\rangle_{S_{\mbox{\tiny{\rm V}}}}
&=&\langle g_s\overline{u}\sigma\cdot {\cal G} u\rangle_{\mbox{\tiny
{\rm 0}}}-\chi_{\rm m}
S_{\mbox{\tiny{\rm V}}}\langle g_s\overline{u}\sigma\cdot {\cal G}
u\rangle_{\mbox
{\tiny{\rm 0}}}\ ,
\label{uqc}
\\*[7.2pt]
\langle g_s\overline{d}\sigma\cdot {\cal G} d
\rangle_{S_{\mbox{\tiny{\rm V}}}}
&=&\langle g_s\overline{d}\sigma\cdot {\cal G} d\rangle_{\mbox{\tiny
{\rm 0}}}+\chi_{\rm m}
S_{\mbox{\tiny{\rm V}}}\langle g_s\overline{d}\sigma\cdot {\cal G}
d\rangle_{\mbox
{\tiny{\rm 0}}}\ ,
\label{dqc}
\\*[7.2pt]
\langle g_s\overline{s}\sigma\cdot {\cal G} s
\rangle_{S_{\mbox{\tiny{\rm V}}}}
&=&\langle g_s\overline{s}\sigma\cdot {\cal G} s\rangle_{\mbox{\tiny
{\rm 0}}}-\chi_{\rm ms}
S_{\mbox{\tiny{\rm V}}}\langle g_s\overline{s}\sigma\cdot {\cal G}
s\rangle_{\mbox
{\tiny{\rm 0}}}\ ,
\label{sdc}
\end{eqnarray}
where $\sigma\cdot {\cal G}\equiv
\sigma_{\mu\nu}{\cal G}^{\mu\nu}$ with ${\cal G}^{\mu\nu}$ the gluon
field tensor. One can express $\chi$, $\chi_{\rm s}$, $\chi_{\rm m}$,
and $\chi_{\rm ms}$ in terms of correlation functions
(see Ref.~\cite{jin2}). Here
we have assumed that the response of the up and down quarks is
the same, apart from the sign.
The Wilson coefficients can be calculated following the methods
outlined in Ref.~\cite{jin2}. The results of our calculations for
the invariant functions $\Pi^1_1$ and $\Pi^q_1$
are given in Appendix A.
%


\subsection{Sum rules}

The QCD sum rules are obtained by equating the QCD representation
and the phenomenological representation and applying the Borel
transformation. The resulting sum rules in the proton case
can be expressed as
\begin{eqnarray}
& &{c_1+6c_2\over 2}M^8 E_2 L^{-8/9}
-{c_1+6c_2\over 2}\chi a M^6 E_1
+{3c_2\over 2}\chi_{\rm m}m_0^2 a M^4E_0L^{-14/27}
\nonumber\\*[7.2pt]
& &\hspace*{1.0cm}
+{c_1+3c_2-c_3\over 3} a^2 M^2
=\biggl[2 H_p\,\widetilde{\lambda}_p^2 M_p^2-
\Delta\widetilde{\lambda}_p^2\, M_p M^2
-H_p\,\widetilde{\lambda}_p^2 M^2\biggr]e^{-M_p^2/M^2}
\nonumber\\*[7.2pt]
& &\hspace*{1.5cm}
+\left[
{c_1-6c_2\over 2}s^1_0 a L^{-4/9}
+{3c_2\over 2}m_0^2 a L^{-26/27}\right]\Delta s_0^1 M^2
e^{-s^1_0/M^2}\; ,
\label{sum_p_1}
\\*[14.4pt]
& &-{4c_1-c_3\over 4} a M^4 E_0 L^{-4/9}
-{c_4+c_5-6c_2\over 12}m_0^2 a M^2 L^{-26/27}
+{2c_1\over
3}\chi a^2 M^2 L^{4/9}
\nonumber\\*[7.2pt]
& &\hspace*{0.4cm}
-{c_1+2c_2\over 12}\chi m_0^2 a^2 L^{-2/27}
-{c_1-2c_2\over 12}\chi_{\rm m}m_0^2 a^2L^{-2/27}
\nonumber\\*[7.2pt]
& &\hspace*{0.8cm}=
\biggl[2H_p\,\widetilde{\lambda}_p^2 M_p-
\Delta\widetilde{\lambda}_p^2 M^2\biggr]e^{-M_p^2/M^2}
+{c_3\over 16}(s_0^q)^2\Delta s_0^q
M^2 L^{-8/9} e^{-s^q_0/M^2}\; ,
\label{sum_p_q}
\end{eqnarray}
where $a\equiv -4\pi^2\langle\overline{q}q\rangle_{\mbox{\tiny{\rm 0}}}$,
$\widetilde{\lambda}_p^2\equiv 32\pi^4\lambda_p^2$,
$\Delta\widetilde{\lambda}_p^2\equiv 32\pi^4\Delta\lambda_p^2$,
and $m_0^2\equiv
\langle g_s\overline{q}\sigma\cdot
{\cal G} q\rangle_{\mbox{\tiny{\rm 0}}}/
\langle\overline{q}q\rangle_{\mbox{\tiny{\rm 0}}}$.
Here we have
ignored the isospin breaking in
the vacuum condensates (i.e.,
$\langle\overline{u}\hat{O}u\rangle_{\mbox{\tiny{\rm 0}}}
\simeq\langle\overline{d}\hat{O}d\rangle_{\mbox{\tiny{\rm 0}}}
=\langle\overline{q}\hat{O}q\rangle_{\mbox{\tiny{\rm 0}}}$);
the inclusion of the isospin breaking in vacuum condensates
only gives small refinements of the results.
We have also defined
\begin{eqnarray}
& &E_0\equiv 1-e^{-s^i_0/M^2}\ ,
\nonumber
\\*[7.2pt]
& &E_1\equiv 1-e^{-s^i_0/M^2}\left[{s^i_0\over
M^2}+1\right]\ ,
\nonumber
\\*[7.2pt]
& &E_2\equiv 1-e^{-s^i_0/M^2}\left[{(s^i_0)^2\over 2M^4}
+{s^i_0\over M^2}+1\right]\ ,
\label{conform}
\end{eqnarray}
and
\begin{eqnarray}
& &c_1=(1-t)^2\ ,\hspace{0.8cm}c_2=1-t^2\ ,\hspace{0.8cm}c_3=5t^2+2t+5\ ,
\nonumber
\\*[7.2pt]
& &c_4=t^2+10t+1\ ,\hspace{0.8cm}c_5=t^2+4t+7\ .
\label{c-def}
\end{eqnarray}
The anomalous dimensions
of the various operators have been taken into
account through the factor
$L\equiv\ln(M^2/\Lambda_{\rm QCD}^2)/\ln(\mu^2/\Lambda_{\rm
QCD}^2)$\cite{svz1,ioffe3}. We
take the renormalization scale $\mu$ and the QCD scale parameter
$\Lambda_{\rm QCD}$ to be $500\,\text{MeV}$
 and $150\,\text{MeV}$\cite{ioffe3}.

The sum rules in the $\Sigma^+$ case are given by
\begin{eqnarray}
& &3c_2M^8 E_2L^{-8/9}
-3c_2\chi a M^6 E_1
+{c_1\over 2}\chi_{\rm s}faM^6 E_1
+(c_1-2c_3)m_s a M^4 E_0 L^{-8/9}
\nonumber\\*[7.2pt]
& &\hspace*{0.5cm}
-3c_2m_s f a M^4 E_0 L^{-8/9}
+{3c_2\over 2}\chi_{\rm m}m_0^2 a M^4 E_0 L^{-14/27}
-{c_2\over 4}m_s f_s m_0^2 a M^2L^{-38/27}
\nonumber\\*[7.2pt]
& &\hspace*{0.5cm}
-{2c_1+3c_2-6c_3\over 12}m_s m_0^2 a M^2 L^{-38/27}
+{c_1-2c_3\over 3} f a^2 M^2
+{2c_3\over 3}\chi m_s a^2 M^2
\nonumber\\*[7.2pt]
& &\hspace*{0.5cm}
+c_2\chi m_s f a^2 M^2
+c_2\chi_{\rm s}m_s f a^2 M^2
\nonumber\\*[7.2pt]
& &\hspace*{1.0cm}
=\biggl[2H_{\Sigma^+}
\,\widetilde{\lambda}_{\Sigma^+}^2 M_{\Sigma^+}^2
-\Delta\widetilde{\lambda}_{\Sigma^+}^2\, M_{\Sigma^+} M^2
-H_{\Sigma^+}\,\widetilde{\lambda}_{\Sigma^+}^2 M^2
\biggr]e^{- M_{\Sigma^+}^2/M^2}
+\biggl[{c_1\over 4}m_s (s^1_0)^2L^{-4/3}
\nonumber\\*[7.2pt]
& &\hspace*{1.5cm}
+{c_1\over 2} f a s^1_0 L^{-4/9}
-3c_2a s^1_0 L^{-4/9}
+{3c_2\over 2} m_0^2 a L^{-26/27}\biggr]\Delta s_0^1
M^2  e^{-s^1_0/M^2}\; ,
\label{sum_s_1}
%
%
\\*[14.4pt]
& &3c_2m_sM^6 E_1L^{-4/3}
-{2c_1-c_3\over 2}a M^4 E_0L^{-4/9}
+3c_2 f a M^4 E_0 L^{-4/9}
-3c_2\chi m_s a M^4E_0L^{-4/9}
\nonumber\\*[7.2pt]
& &\hspace*{0.5cm}
-{c_3\over 4}\chi_{\rm s} m_s f a M^4 L^{-4/9}
-{c_2\over 12}m_0^2 a M^2 L^{-26/27}
-{5c_2\over 4}f_s m_0^2 a M^2 L^{-26/27}
\nonumber\\*[7.2pt]
& &\hspace*{0.5cm}
+{7c_2\over 4}\chi_{\rm m}m_s m_0^2 a M^2 L^{-26/27}
-{c_5\over 12}\chi_{\rm ms} m_s f_s m_0^2 a M^2
L^{-2/27}
+{2c_1\over 3}\chi a^2 M^2 L^{4/9}
\nonumber\\*[7.2pt]
& &\hspace*{0.5cm}
-2c_2\chi f a^2 M^2 L^{4/9}
-2c_2\chi_{\rm s} f a^2 M^2 L^{4/9}
-c_2m_s a^2 L^{-4/9}
-{c_3-2c_1\over 6} m_s f a^2 L^{-4/9}
\nonumber\\*[7.2pt]
& &\hspace*{0.5cm}
-{c_1\over 12}\chi m_0^2 a^2 L^{-2/27}
+{5c_2\over 12}\chi f_s m_0^2 a^2 L^{-2/27}
+{7c_2\over 12}\chi_{\rm s}f m_0^2 a^2 L^{-2/27}
\nonumber\\*[7.2pt]
& &\hspace*{0.5cm}
-{c_1\over 12}\chi_{\rm m} m_0^2 a^2 L^{-2/27}
+{5c_2\over 12}\chi_{\rm ms} f_s m_0^2 a^2 L^{-2/27}
+{7c_2\over 12}\chi_{\rm m} f m_0^2 a^2 L^{-2/27}
\nonumber\\*[7.2pt]
& &\hspace*{1.0cm}
=
\biggl[2H_{\Sigma^+}\,\widetilde{\lambda}_{\Sigma^+}^2 M_{\Sigma^+}-
\Delta\widetilde{\lambda}_{\Sigma^+}^2 M^2\biggr]e^{-M_{\Sigma^+}^2/M^2}
\nonumber\\*[7.2pt]
& &\hspace*{1.5cm}
+\biggl[{c_3\over 16}(s^q_0)^2-3c_2m_s a
-{c_3\over 4}m_s f a\biggr]\Delta s_0^q
M^2 L^{-8/9} e^{-s^q_0/M^2}\; ,
\label{sum_s_q}
\end{eqnarray}
where $f\equiv \langle\overline{s}s\rangle_{\mbox{\tiny{\rm 0}}}
/\langle\overline{q}q\rangle_{\mbox{\tiny{\rm 0}}}$ and
$f_s\equiv
\langle g_s\overline{s}\sigma\cdot
{\cal G} s\rangle_{\mbox{\tiny{\rm 0}}}/
\langle g_s\overline{q}\sigma\cdot
{\cal G} q\rangle_{\mbox{\tiny{\rm 0}}}$.
 The sum rules in the $\Xi^0$ case are
\begin{eqnarray}
& &-{c_1\over 2}M^8 E_2 L^{-8/9}
+{c_1\over 2}\chi a M^6 E_1
-3c_2\chi_{\rm s}f a M^6 E_1
-3c_2m_s a M^4 E_0 L^{-8/9}
\nonumber\\*[7.2pt]
& &\hspace*{0.5cm}
+(c_1-2c_3)m_s f a M^4 E_0 L^{-8/9}
+{3c_1\over 2}\chi_{\rm ms}
f_s m_0^2 a M^4 E_0 L^{-14/27}
\nonumber\\*[7.2pt]
& &\hspace*{0.5cm}
-{c_2\over 4}m_s m_0^2 a M^2 L^{-38/27}
-{2c_1+3c_2-6c_3\over 12}m_s f_s m_0^2 a M^2 L^{-38/27}
-{c_3\over 3}f^2 a^2 M^2
\nonumber\\*[7.2pt]
& &\hspace*{0.5cm}
-c_2f a^2 M^2
-(c_1-2c_3)\chi m_s f a^2 M^2
-(c_1-2c_3)\chi_{\rm s}m_s f a^2 M^2
\nonumber\\*[7.2pt]
& &\hspace*{1.0cm}
=\biggl[2H_{\Xi^0}\,\widetilde{\lambda}_{\Xi^0}^2
M_{\Xi^0}^2
-\Delta\widetilde{\lambda}_{\Xi^0}^2\, M_{\Xi^0} M^2
-H_{\Xi^0}\,\widetilde{\lambda}_{\Xi^0}^2 M^2\biggr]
e^{-M_{\Xi^0}^2/M^2}
\nonumber\\*[7.2pt]
& &\hspace*{2.0cm}
+
\biggl[-{3c_2\over 2}m_s (s^1_0)^2 L^{-4/3}
+{c_1\over 2}s^1_0 a L^{-4/9}
\nonumber\\*[7.2pt]
& &\hspace*{1.5cm}
-3c_2s_0 f a L^{-4/9}
+{3c_2\over 2}f_s m_0^2 a L^{-26/27}\biggr]\Delta s_0^1
e^{-s^1_0/M^2}\; ,
\label{sum_x_1}
%
\\*[14.4pt]
& &3c_2m_s M^6 E_1 L^{-4/3}
+{c_3\over 4} a M^4 E_0 L^{-4/9}
+3c_2f a M^4 E_0 L^{-4/9}
-3c_2\chi m_s a M^4 E_0 L^{-4/9}
\nonumber\\*[7.2pt]
& &\hspace*{0.5cm}
+{2c_1-c_3\over 2}\chi_{\rm s}m_s f a M^4 E_0 L^{-4/9}
+{c_5\over 12}m_0^2 a M^2 L^{-26/27}
-{7c_2\over 4}f_s m_0^2 a M^2 L^{-26/27}
\nonumber\\*[7.2pt]
& &\hspace*{0.5cm}
+{5c_2\over 4}\chi_{\rm m}m_s m_0^2 a M^2 L^{-14/27}
+{c_4\over 12}\chi_{\rm ms}m_s f_s m_0^2 M^2 L^{-26/27}
-2c_2\chi f a^2 M^2 L^{4/9}
\nonumber\\*[7.2pt]
& &\hspace*{0.5cm}
-2c_2\chi_{\rm s} f a^2 M^2 L^{4/9}
+{2c_1\over 3}\chi_{\rm s} f^2 a^2 M^2 L^{4/9}
-c_2m_s f^2 a^2 L^{-4/9}
-{c_3-2c_1\over 6}m_s f_s a^2 L^{-4/9}
\nonumber\\*[7.2pt]
& &\hspace*{0.5cm}
+{7c_2\over 12}\chi f_s m_0^2 a^2 L^{-2/27}
-{c_1\over 12}\chi_{\rm s}f f_s m_0^2 a^2L^{-2/27}
+{5c_2\over 12}\chi_{\rm s}f m_0^2 a^2 L^{-2/27}
\nonumber\\*[7.2pt]
& &\hspace*{0.5cm}
+{5c_2\over 12}\chi_{\rm m}f m_0^2 a^2 L^{-2/27}
-{c_1\over 12}\chi_{\rm ms}f f_s m_0^2 a^2  L^{-2/27}
+{7c_2\over 12}\chi_{\rm ms}f_s m_0^2 a^2 L^{-2/27}
\nonumber\\*[7.2pt]
& &\hspace*{1.0cm}=
\biggl[2H_{\Xi^0}\,\widetilde{\lambda}_{\Xi^0}^2 M_{\Xi^0}-
\Delta\widetilde{\lambda}_{\Xi^0}^2\biggr]e^{-M_{\Xi^0}^2/M^2}
\nonumber\\*[7.2pt]
& &\hspace*{2.0cm}
+\biggl[
{c_3\over 16}(s^q_0)^2
-3c_2m_s a
-{c_3-2c_1\over 2} m_s f a\biggr]\Delta s_0^q
 L^{-8/9} M^2 e^{-s_0^q/M^2}\; .
\label{sum_x_q}
\end{eqnarray}
{}.


\section{Sum-rule analysis}
\label{anay}

We now analyze the sum rules derived in the previous section
and extract the baryon matrix elements of interest. Here
we follow Ref.~\cite{jin2} and use only the
sum rules Eqs.~(\ref{sum_p_q}), (\ref{sum_s_q}), and
(\ref{sum_x_q}), which are more stable than the other
three sum rules. The pattern that one of the sum rules
(in each case) works well while the other does not
has been seen in various external field
problems\cite{ioffe1,chiu1,chiu2,jin1,jin2}.
This may be attributed to the different asymptotic
behavior of various sum rules. As emphasized earlier,
the phenomenological side of the external field sum rules
contains single pole terms arising from the transition
between the ground state and the excited states, whose
contribution is {\it not} suppressed relative to the
double pole term and thus contaminates
the double pole contribution. The degree of this
contamination may vary from one sum rule to another.
The sum rule with smaller single pole contribution
works better.
We refer the reader to Refs.~\cite{chiu2,jin1,jin2}
for more discussion
about the different behavior of various external field
sum rules.
In the analysis to follow, we disregard the sum
rule Eqs.~(\ref{sum_p_1}), (\ref{sum_s_1}), and
(\ref{sum_x_1}), and consider only the results from
the sum rules  Eqs.~(\ref{sum_p_q}), (\ref{sum_s_q}), and
(\ref{sum_x_q}).

We adopt the numerical optimization procedures used in
Refs.~\cite{leinweber2,furnstahl1}. The
sum rules are sampled in the fiducial region of Borel $M^2$, where
the contributions from the high-dimensional condensates
remain small and the continuum contribution is controllable.
We choose
\begin{eqnarray}
& &
0.8\leq M^2\leq 1.4\, {\mbox{GeV}}^2\hspace*{2cm}
{\mbox{for proton case}}\ ,
\\*[7.2pt]
& &
1.2\leq M^2\leq 1.8\, {\mbox{GeV}}^2\hspace*{2cm}
{\mbox{for}}\,\Sigma^+\,{\mbox{and}}\,\Xi^0 {\mbox{case}}\ ,
\end{eqnarray}
which have been identified as the fiducial region for the baryon
mass sum rules\cite{ioffe1,ioffe5}. Here we adopt these boundaries as
the maximal limits of applicability of the external field sum
rules. The sum-rule predictions are obtained by
minimizing the logarithmic measure
$\delta (M^2)={\mbox{ln}}[{\mbox{maximum}}\{{\mbox{LHS,RHS}}\}/
{\mbox{minimum}}\{{\mbox{LHS,RHS}}\}]$ averaged over $150$ points
evenly spaced within the fiducial region of $M^2$, where
LHS and RHS denote the left- and right-hand sides of
the sum rules, respectively.

Note that the {\it vacuum} spectral parameters $\lambda_{\rm B}^2$,
$M_{\rm B}$ and $s_0^i$, also appear in the external field sum rules
 Eqs.~(\ref{sum_p_1}--\ref{sum_p_q}) and
(\ref{sum_s_1}--\ref{sum_x_q}).
Here we use the experimental values for
the baryon masses and extract $\lambda_{\rm B}^2$
and $s_0^i$ from baryon mass sum rules using the same
optimization procedure as described above.
We then extract $H_{\rm B}$, $\Delta\lambda_{\rm B}^2$, and
$\Delta s_0^i$ from the external field sum rules.

For vacuum condensates, we use $a=0.55\, {\mbox{GeV}}^3\, (m_u
+m_d\simeq 11.8{\mbox{MeV}})$\cite{ioffe1,ioffe3},
$m_0^2=0.8\, {\mbox{GeV}}^2$\cite{ioffe1,belyaev3},
and $f\simeq f_s=0.8$\cite{belyaev3,leinweber2}.
We take the strange quark mass
$m_s$ to be $150\,  {\mbox{MeV}}$\cite{ioffe5}. The
parameter $\chi$ has been estimated in Ref.~\cite{jin2}.
The estimate in chiral perturbation theory gives
$\chi\simeq 2.2\, {\mbox{GeV}}^{-1}$. It is also shown that
to the lowest order in $\delta m$, $\chi$ is determined by
\begin{equation}
\chi\delta m=-\gamma+O[(\delta m)^2]\ ,
\label{chi-est}
\end{equation}
where $\gamma\equiv \langle\overline{d}d
\rangle_{\mbox{\tiny{\rm 0}}}/\langle\overline{u}u
\rangle_{\mbox{\tiny{\rm 0}}}-1$, and $\delta m$
has been determined  by Gasser and Leutwyler,
$\delta m /(m_u + m_d ) = 0.28 \pm 0.03$\cite{gasser1}.
The value of $\gamma$ has been estimated previously in various
approaches\cite{gasser2,paver1,pascual1,bagan1,dominguez1,%
dominguez2,narison1,adami2,adami1,eletsky1}
 with results ranging from $-1\times 10^{-2}$
to $-2\times 10^{-3}$, which upon using Eq.~(\ref{chi-est})
and a median value for $\delta m=3.3\,\text{MeV}$,
corresponds to
\begin{equation}
0.5\,\text{GeV}^{-1}\leq\chi\leq 3.0 \,\text{GeV}^{-1}\ .
\label{chi-range}
\end{equation}
We shall consider this range of $\chi$ values.
We follow Ref.~\cite{jin2} and assume
$\chi_{\rm m}\simeq \chi$,
which is equivalent to the assumption that $m_0^2$
is isospin independent.

The parameter $\chi_{\rm s}$ measures the response of
the strange quark condensate to the external field, which
has not been estimated previously.
Since $\overline{s}s$ is an isospin scalar operator,
$\chi_{\rm s}$ arises from the isospin mixing and
we expect $\chi_{\rm s}<\chi$.
Following Ref.~\cite{jin2},
one may express $\chi_{\rm s}$ in terms of a
correlation function
and estimate it in chiral perturbation theory. It is
easy to show that $\chi_{\rm s}\langle\overline{s}s
\rangle_{\mbox{\tiny{\rm 0}}}={d\over d\delta m}
\langle\overline{s}s
\rangle_{\mbox{\tiny{\rm 0}}}$. So, one may determine
$\chi_{\rm s}$ by evaluating ${d\over d\delta m}
\langle\overline{s}s
\rangle_{\mbox{\tiny{\rm 0}}}$ in effective QCD models.
Here we shall treat $\chi_{\rm s}$ as a free parameter
and consider the values of $\chi_{\rm s}$ in the
range of $0\leq\chi_{\rm s}\leq 3.0\,\text{GeV}^{-1}$.
We also assume that $\chi_{\rm ms}\simeq \chi_{\rm s}$.

We first analyze the sum rules for Ioffe's
interpolating field (i.e., $t=-1$). We start from
the proton case. The optimized result for $H_p$
as function of $\chi$ is plotted
in Fig.~\ref{fig-1}. One can see that
$H_p$ varies rapidly with $\chi$. Therefore,
the sum-rule prediction for the proton matrix element
$H_p$ depends strongly on the response of the up and down
quark condensates to the external source.
(The sum rules in the proton case
are independent of $\chi_{\rm s}$ and $\chi_{\rm sm}$.)
For moderate values of $\chi$ ($1.5\,\text{GeV}^{-1}
\leq\chi\leq 2.0\,\text{GeV}^{-1}$), the predictions
are
\begin{equation}
H_p\simeq 0.54-0.78\ .
\label{typ-p}
\end{equation}
On the other hand, for large values of
$\chi$ ($2.4\,\text{GeV}^{-1}\leq\chi
\leq 3.0\,\text{GeV}^{-1}$), we find $H_p\simeq
0.97-1.25$. For small values of $\chi$
($\chi\leq 1.4\,\text{GeV}^{-1}$),
the continuum contribution is larger than $50\%$,
implying that the continuum contribution is dominant
in the Borel region of interest and the prediction
is not reliable.
 The predictions for $\Delta
\lambda_p^2$ and $\Delta s_0^q$ also change with $\chi$
 in the same way as $H_p$.

To see how well the sum rule works, we plot the LHS, RHS,
and the individual terms of RHS of Eq.~(\ref{sum_p_q}) as functions
of $M^2$ with $\chi=1.8\,\text{GeV}^{-1}$ in Fig.~\ref{fig-2}
using the optimized values for $H_p$, $\Delta\lambda_p^2$,
and $\Delta s_0^q$. We see that the solid (LHS) and long-dashed (RHS)
curves are right on top of each other, showing a very good
overlap. We also note from Fig.~\ref{fig-2}
that the first term of RHS (curve 1) is larger than
the second (curve 2) and third (curve 3) terms. This shows that
the double pole contribution is stronger than the single
pole contribution and the predictions are thus stable.
(Although the second and third terms are sizable
individually, their sum is small.)

In Fig.~\ref{fig-3}, we have displayed the predicted
$H_{\Sigma^+}$ as function of $\chi$ for three
different values of $\chi_{\rm s}$. One notices that
$H_{\Sigma^+}$ is largely insensitive to $\chi_{\rm s}$,
but strongly dependent on $\chi$ value. For $\chi$
values in the range of $2.2\,\text{GeV}^{-1}\leq
\chi\leq 3.0\,\text{GeV}^{-1}$, we find
\begin{equation}
H_{\Sigma^+}\simeq 1.65-2.48\ .
\label{typ-s}
\end{equation}
For smaller $\chi$, we obtain smaller values for $H_{\Sigma^+}$.
The predictions for $\Delta\lambda_{\Sigma^+}^2$
and $\Delta s_0^q$ change in a similar pattern. The sum rule
works very well and the continuum contribution is small for
all $\chi$ and $\chi_{\rm s}$ values considered here.

The optimized $H_{\Xi^0}$ as function of $\chi_{\rm s}$
is shown in Fig.~\ref{fig-4}. [When $t=-1$, the sum rule
Eq.~(\ref{sum_x_q}) is independent of $\chi$
and $\chi_{\rm s}$.]  We see that
the result is very sensitive to the $\chi_{\rm s}$ value.
Thus the prediction for $H_{\Xi^0}$ has a strong dependence
on the response of the strange quark condensate to the
external field.
For moderate $\chi_{\rm s}$ ($1.7\,\text{GeV}^{-1}\leq
2.2\,\text{GeV}^{-1}$), we get
\begin{equation}
H_{\Xi^0}\simeq 1.57-1.84\ .
\label{typ-x}
\end{equation}
For larger (smaller) values of $\chi_{\rm s}$, we find
larger (smaller) values for $H_{\Xi^0}$.
At $\chi_{\rm s}=0$, we get
$H_{\Xi^0}\simeq 0.68$. The results for
$\Delta\lambda^2_{\Xi^0}$ and $\Delta s_0^q$ increase
(decrease) as $\chi_{\rm s}$ increases (decreases).

All of the results above use Ioffe's interpolating field
(i.e., $t=-1$); we now present the results for general
interpolating field. In Fig.~\ref{fig-5}, we have plotted
the predicted $H_p$, $H_{\Sigma^+}$, and $H_{\Xi^0}$ as
functions of $t$ for $\chi=2.5\,\text{GeV}^{-1}$ and
$\chi_{\rm s}=1.5\,\text{GeV}^{-1}$. As $t$ increases,
$H_p$, $H_{\Sigma^+}$, and $H_{\Xi^0}$ all increase; the
rate of increase is essentially the same for $H_p$
and $H_{\Sigma^+}$, but somewhat smaller for $H_{\Xi^0}$.
We note that the {\it vacuum} spectral parameters $\lambda^2_{\rm B}$
and $s^q_0$ decrease as $t$ increases; this leads to
a large variation of $H_p$, $H_{\Sigma^+}$, and $H_{\Xi^0}$
with $t$.

The sensitivity of our results to the assumption of
$\chi_{\rm m}=\chi$ is displayed in Fig.~\ref{fig-6},
where $t$ and $\chi_{\rm s} (=\chi_{\rm ms})$ are
fixed at $-1$ and $1.5\,\text{GeV}^{-1}$, respectively.
The three curves are obtained by using $\chi_{\rm m}
=\chi$, ${1\over 2}\chi$, and ${3\over 2}\chi$, respectively.
We note that $H_p$ and $H_{\Sigma^+}$
get larger (smaller) as $\chi_{\rm m}$ becomes
smaller (larger). The results are more sensitive
to $\chi_{\rm m}$ in the proton case than in the $\Sigma^+$
case. The prediction for $H_p$ changes by about $25\%$
while the prediction for $H_{\Sigma^+}$ changes
by about $15\%$ when the $\chi_{\rm m}$ value
is changed by $50\%$. This implies that the terms
proportional to $\chi_{\rm m}$ in the sum rules
give rise to sizable contributions. The sensitivity
of our predictions to the assumption of
$\chi_{\rm sm}=\chi_{\rm s}$ is illustrated in
Fig.~\ref{fig-7}, with $t=-1$ and $\chi=\chi_{\rm m}
=2.5\,\text{GeV}^{-1}$. The three curves correspond
to $\chi_{\rm ms}=\chi_{\rm s}$, ${1\over 2}\chi_{\rm s}$,
and ${3\over 2}\chi_{\rm s}$, respectively. One can see that both
$H_{\Sigma^+}$ and $H_{\Xi^0}$ are insensitive to
changes in $\chi_{\rm ms}$. This indicates that
the terms proportional to $\chi_{\rm ms}$ give
only small contributions to the sum rules. One
also notices that $H_{\Sigma^+}$ depends only
weakly on $\chi_{\rm s}$. Finally, the effect of
ignoring the response of continuum threshold is
shown in Fig.~\ref{fig-8}. The solid (dashed) curve
is obtained by including (omitting) the third term
on the RHS of Eq.~(\ref{sum_p_q}). The difference
between the two curves is large for moderate and
large values of $\chi$. This shows that the response
of the continuum threshold  can be sizable and
should be included in the sum rules. Unfortunately,
the response
of the continuum thresholds has been omitted
in all previous works on external field sum rules.
This was first noticed by Ioffe\cite{ioffe2}.

\section{Estimate of baryon isospin mass splittings}
\label{isospin}

In this section we estimate the baryon isospin mass splittings
using $\delta m$ and the baryon matrix elements of
isovector-scalar current calculated in the previous section.


The observed hadron isospin mass splittings arise from electromagnetic
interaction and from the difference between up and down quark masses:
\begin{equation}
\delta m_h=(\delta m_h)_{\rm el}+(\delta m_h)_{\rm q}\ ,
\label{dm-sep}
\end{equation}
where $(\delta m_h)_{\rm el}$ and $(\delta m_h)_{\rm q}$ denote the
contributions due to electromagnetic interaction and due to
the up and down quark mass difference, respectively.\footnote%
{This separation is renormalization scale dependent. However,
this scale dependence is weak; it is thus meaningful to
separate the contribution of quark mass difference
from that due to electromagnetic interaction (see Ref.~\cite{jin2}).}
Following Ref.~\cite{jin2}, one can treat $\delta m$
as a small parameter and  using the  Hellman-Feynman
theorem~\cite{hellman1,feynman1} to show that
the octet baryon
isospin mass splittings to first order in $\delta m$
can be expressed as
\begin{eqnarray}
& &M_n-M_p=(M_n-M_p)_{\rm el}+\delta m H_p\ ,
\label{mq-np}
\\*[7.2pt]
& &M_{\Sigma^-}-M_{\Sigma^+}=
(M_{\Sigma^-}-M_{\Sigma^+})_{\rm el}+\delta m H_{\Sigma^+}\ ,
\label{mq-sig}
\\*[7.2pt]
& &M_{\Xi^-}-M_{\Xi^0}=(M_{\Xi^-}-M_{\Xi^0})_{\rm el}
+\delta m H_{\Xi^0}\ .
\label{mq-xi}
\end{eqnarray}
Note that $H_n=-H_p$, $H_{\Sigma^-}=-H_{\Sigma^+}$,
and $H_{\Xi^-}=-H_{\Xi^0}$ to the lowest order
in $\delta m$.
%
%
Therefore, QCD sum rule predictions for $H_p$, $H_{\Sigma^+}$,
and $H_{\Xi^0}$,
along with the electromagnetic contributions\cite{gasser1}
\begin{eqnarray}
& &(M_n-M_p)_{\rm el}= -0.76\pm 0.30\,{\mbox{MeV}}\ ,
\label{mel-np}
\\*[7.2pt]
& &(M_{\Sigma^-}-M_{\Sigma^+})_{\rm el}= 0.17\pm0.3\,{\mbox{MeV}}\ ,
\label{mel-sig}
\\*[7.2pt]
& &(M_{\Xi^-}-M_{\Xi^0})_{\rm el}=0.86\pm 0.30\,{\mbox{MeV}}\ ,
\label{mel-xi}
\end{eqnarray}
will lead to an estimate of the baryon isospin mass splittings.
Taking the experimental mass difference\cite{particle1}, one finds
\begin{eqnarray}
& &(M_n-M_p)_{\rm q}^{\rm exp}=2.05\pm 0.30\,\text{MeV}\ ,
\\*[7.2pt]
& &(M_{\Sigma^-}-M_{\Sigma^+})_{\rm q}^{\rm exp}=7.9\pm 0.33\,\text{MeV}\ ,
\\*[7.2pt]
& &(M_{\Xi^-}-M_{\Xi^0})_{\rm q}^{\rm exp}=5.54\pm 0.67\,\text{MeV}\ .
\label{mass-dif-exp}
\end{eqnarray}

We have seen from last section that the uncertainties in our
knowledge of the response of the quark condensates to the
external field, $\chi$ and $\chi_{\rm s}$, leads to uncertainties
in the sum-rule determination of the baryon matrix elements
$H_{\rm B}$. (There are also uncertainties in $\delta m$.)
Therefore, our estimate here are only {\it qualitative}.
For most of the values for $t$, $\chi$ and $\chi_{\rm s}$
considered here, the sum-rule analysis gives
$0 < H_p < H_{\Xi^0}\leq H_{\Sigma^+}$ (see Figs.~\ref{fig-5}, \ref{fig-6}
and \ref{fig-7}), which implies
\begin{equation}
0<(M_n-M_p)_{\rm q}<(M_{\Xi^-}-M_{\Xi^0})_{\rm q}\leq
(M_{\Sigma^-}-M_{\Sigma^+})_{\rm q}\ .
\label{quli}
\end{equation}
This qualitative feature is compatible with the experimental data.
For the baryon interpolating fields  with $t=-1$
and moderate $\chi$ and $\chi_{\rm s}$ values
($1.6\,\text{GeV}^{-1}\leq\chi\leq 2.2\,\text{GeV}^{-1}$
and $1.3\,\text{GeV}^{-1}\leq\chi_{\rm s}\leq 1.8\,\text{GeV}^{-1}$),
we get
\begin{eqnarray}
& &1.95\,\text{MeV}\leq (M_n-M_p)_{\rm q}\leq 2.41\,\text{MeV}\ ,
\\*[7.2pt]
& &4.0\,\text{MeV}\leq (M_{\Sigma^-}-M_{\Sigma^+})_{\rm q}\leq 6.3\,\text{MeV}\
,
\\*[7.2pt]
& &4.5\,\text{MeV}\leq (M_{\Xi^-}-M_{\Xi^0})_{\rm q}\leq 5.38 \,\text{MeV}\ ,
\label{mass-est}
\end{eqnarray}
where we have used a median value
$\delta m\simeq 3.3\,\text{MeV}$. These results are
comparable to the experimental data, though the
result in the $\Sigma$ case is somewhat too small.
Smaller and larger values of $\chi$ and $\chi_{\rm s}$
lead to correspondingly smaller and larger values for
the baryon isospin mass differences. As $t$ increases
(decreases), the results increase (decrease).

\section{Discussion}
\label{discussion}

Our primary goal in the present paper has been to extract the
baryon matrix element $H_{\rm B}=\langle B|\overline{u}u-
\overline{d}d|B\rangle/2M_{\rm B}$ for octet baryons. We observe that
the sum-rule predictions for $H_{\rm B}$ are quite sensitive
to the response of quark condensates to the external
isovector-scalar field, which is not well determined.
This means that our conclusion about $H_{\rm B}$ can only be
{\it qualitative} at this point. The most concrete conclusion
we can draw from this work is that QCD sum rules
predict positive values for $H_p$, $H_{\Sigma^+}$,
and $H_{\Xi^0}$ and $H_p<H_{\Xi^0}\leq H_{\Sigma^+}$.
This qualitative feature is, for the most part,
stable against variations of the response of the condensates
to the external source and the choice of baryon interpolating
fields.

We note that the inequality $H_p<H_{\Xi^0}\leq H_{\Sigma^+}$ indicates
SU(3) symmetry violation in the baryon matrix elements of the
isovector-scalar current. This arises mainly from the difference
in the baryon interpolating fields used in the QCD sum rules
and from the fact that the
isovector-scalar current is not a SU(3) singlet. Clearly, it is
a very interesting topic to check this inequality in other
effective QCD models. At this stage,
it is unclear whether the difference in the baryon interpolating fields is
connected to the SU(3) symmetry breaking in the baryon wave functions.

In the present study, we derived and used a complete form for the
phenomenological representation, which has also been given in
Ref.~\cite{ioffe2}. This form includes the response
of the continuum thresholds, which was ignored in
Ref.~\cite{jin2}. We found that
the neglect of the response of the continuum thresholds
can have large effect on the extraction of the baryon matrix
elements. This suggests that the contribution arising from the
response of the continuum thresholds, neglected in previous works,
should be accounted in the study of general
external field sum rules
(see Ref.~\cite{ioffe2} for estimates of the effects
of this contribution on the extraction of various physical
quantities).

The spectral parameters in the absence of the external
source, $M_{\rm B}$, $\lambda^2_{\rm B}$, and $s_0^i$, appear
in all external field sum rules. Unlike the mass, there are
no experimental values for the coupling $\lambda^2_{\rm B}$
and the thresholds $s_0^i$. One usually evaluates these
parameters from the mass sum rules by fixing the mass at the
experimental value.  This means that the uncertainties
associated with the vacuum spectral parameters will
give rise to additional uncertainties in the determination of the
baryon matrix elements of various current, besides the uncertainties
in the external field sum rules themselves. This is a general drawback
of the external field sum-rule approach. It is also worth
pointing out that it is
the product of $\lambda^2_{\rm B}$ and the baryon
matrix element appears in the external field sum rules
[see Eqs.~(\ref{sum_p_1}--\ref{sum_x_q})].
So, it is more suitable to determine the product
of $\lambda^2_{\rm B}$ and the baryon
matrix element from the external field sum rules; one then needs
a good knowledge of $\lambda^2_{\rm B}$ in order to
extract the baryon matrix element cleanly.

The sum-rule predictions are fairly sensitive to the choice
of baryon interpolating fields. This sensitivity arises
from {\it both} the dependence of the truncated OPE result
{\it and} the dependence of the extracted parameters
$\lambda^2_{\rm B}$ and $s_0^i$ on the choice of the
baryon interpolating fields. We found that the latter has
stronger dependence, and hence leads to larger contribution
to the change of the predictions with $t$.

The non-electromagnetic part of the baryon isospin
mass difference is essentially given by the matrix element
$H_{\rm B}$ multiplied by the light quark mass difference $\delta m$.
Given the uncertainties in the determination of $H_{\rm B}$
mentioned above, our estimate of the isospin mass splittings
for the octet baryons must be qualitative. It is found that
the QCD sum-rule predictions yield $(M_n-M_p)_{\rm q}<
(M_{\Xi^-}-M_{\Xi^0})_{\rm q}\leq (M_{\Sigma^-}-M_{\Sigma^+})
_{\rm q}$. This qualitative result is consistent with the
experimental data and insensitive to the details of
calculation. If we use a
median value $\delta m=3.3\,\text{MeV}$ and moderate
values for $\chi$ and $\chi_{\rm s}$, we obtain results
comparable to the experimental values. However, since
the response of various condensates to the external source and
$\delta m$ are not precisely known and the uncertainties
from other sources cannot be accessed systematically,
it is not wise
to make a critical comparison with data or to attempt
to extract $\chi$ [and hence $\gamma$ through Eq.~(\ref{chi-est})]
and $\chi_{\rm s}$ by fitting the experimental data.
Clearly, further study of the response of the quark
condensates to external isovector-scalar field is
important, along with more accurate determination
of the vacuum spectral parameters. Effective QCD
models may give some independent information on the
response of the quark condensates while the lattice
QCD may offer clean determination of the vacuum
spectral parameters\cite{leinweber1}.

There have been several earlier papers that study
the neutron-proton mass difference\cite{pascual1,adami2,%
adami1,hatsuda1,hatsuda2,yang1} and the
baryon isospin mass splittings for other octet
baryons\cite{pascual1,adami1}, based on QCD sum-rule
approach. In Ref.~\cite{pascual1}, the baryon mass
differences were extracted directly from the baryon
mass sum rules by including the quark mass difference and
the isospin breaking in the quark condensates. The
contributions of quark-gluon mixed condensates
were ignored, and somewhat different values for the
vacuum condensates and the strange quark mass were
used. This can lead to large effects on the extraction of
the isospin mass splittings. The procedure for
analyzing the sum rules was also quite different
from the one used in the present paper.  In
Ref.~\cite{hatsuda1,adami2}, the neutron-proton mass was extracted
from the difference between the neutron and proton mass sum rules, but
the continuum contributions were disregarded. In a later calculation
\cite{hatsuda2}, the authors of Ref.~\cite{hatsuda1} have included the
continuum contribution in the study of the density dependence of the
neutron-proton mass difference in the medium.
In these works, the contributions from the quark-gluon
condensates and the change in the continuum thresholds
were omitted.
The study of neutron-proton mass difference in Ref.~\cite{yang1}
was based on the mass sum rules directly.
Apart from keeping the quark mass difference and the quark
condensates difference, an attempt was made to incorporate the
electromagnetic contribution also phenomenologically in the sum rules.

The analysis in Ref.~\cite{adami1} is more closely related
to the present work. The goal of Ref.~\cite{adami1} was,
however, to determine the parameters $\delta m$ and
 $\gamma$ by fitting
all isospin mass splittings in the baryon octet. The sum
rules were obtained for Ioffe's interpolating field by
treating the quark mass and the isospin breaking in quark
condensates as perturbations. On the phenomenological sides
of the sum rules, all spectral parameters, mass, residue,
and continuum thresholds, were allowed to change. Note that
the sum rules derived by us in Sec.~\ref{sumrule} can also be derived
directly from the mass sum rules.  Writing $m_u =\hat m - \delta m /2$,
$m_d =\hat m + \delta m /2$ and assuming
$\chi = -\gamma /\delta m$ [see Eq.~(\ref{chi-est})], one
can differentiate the mass sum rules with respect to $\delta m$.
For $t=-1$, one can
then identify our sum rules Eqs.~(\ref{sum_p_1}--\ref{sum_p_q}) and
(\ref{sum_s_1}--\ref{sum_x_q}) with the sum rules given in Ref.~\cite{adami1}.
This coincidence between the sum rules is not surprising, since the quark
mass term in the QCD Lagrangian can also be regarded as a constant
external scalar field.
We observe, however, that the contributions from
dimension eight condensates have not been included in
Ref.~\cite{adami1}. We have seen in our analysis of the sum
rules (see Figs.~\ref{fig-6} and \ref{fig-7}) that
these contributions can be numerically significant.
In addition, the
authors of Ref.~\cite{adami1} directly used the $\Sigma$ and $\Xi$ mass
sum rules from Ref.~\cite{belyaev3}, where all the terms proportional to
$m_u$ or $m_d$ were neglected. Consequently, some terms proportional
to $m_s$ were not taken into account in the $\Sigma$ and $\Xi$
cases and there was a factor two omitted in the contribution
from four-quark condensates in the nucleon case.

We note that the authors of Ref.~\cite{adami1}
took a very different procedure in analyzing the sum rules.
They used both sum rules to eliminate $\Delta\lambda^2_{\rm B}$
while we used only the more stable one. The continuum
contribution in sum rules
Eqs.~(\ref{sum_p_1}), (\ref{sum_s_1}), and (\ref{sum_x_1})
is large. So, these sum rules are likely to be dominated
by the single pole terms and the predictions based on these
sum rules may not be reliable. The size of the continuum
contribution was not checked in  Ref.~\cite{adami1}.
Certain assumptions such as $\Delta s_0^i=0$ were also used
in some cases. We notice that the absence of continuum contribution
in the external field sum rules does not necessarily
imply $\Delta s_0^i=0$. In fact, as long as
there is continuum contribution in the mass sum rules, one must
include $\Delta s_0^i$ as a unknown quantity to be determined
from the sum rules. Any assumption about $\Delta s_0^i$
may bypass the information extracted for other quantities.
The authors of Ref.\cite{adami1} claimed that
consistency of the two sum rules can be achived for $\gamma=-(2\pm 1)
\times 10^{-3}$, which is different from the values discussed in the
present paper (see discussions in Sec.~\ref{anay}). This discrepancy
arises mainly from the difference in the procedures for analyzing
the sum rules.


\acknowledgements

This work was supported
in part by the Natural Sciences and Engineering
Research Council of Canada.
\newpage
\appendix
\label{app}
\section*{A}

In this appendix, we give the OPE results for the
invariant functions $\Pi^1_1$ and $\Pi^q_1$.
The interpolating fields defined in Eqs.~(\ref
{eta-p}--\ref{eta-x}) are used in the calculation.
We work to the leading order in the perturbation
theory and to the first order in the strange quark
mass $m_s$. The contributions proportional to
up and down quark masses and to gluon
condensates are neglected. Condensates up to
dimension eight are considered.

\begin{eqnarray}
&{\mbox{proton}}:\hspace{0.7cm}&\Pi^1_1(q^2)={c_1+6c_2\over 128
 \pi^4}(q^2)^2\ln(-q^2)+{c_1+6c_2\over
16\pi^2}\chi\langle\overline{q}q\rangle_{\mbox{\tiny{\rm 0}}}
q^2\ln(-q^2)
\nonumber
\\*[7.2pt]
& &\hspace*{1.8cm}
-{3c_2\over 16\pi^2}\chi_{\mbox{\tiny m}}
\langle g_s\overline{q}\sigma\cdot {\cal G}q
\rangle_{\mbox{\tiny{\rm 0}}}\ln(-q^2)
+{c_1+3c_2-c_3\over 6}
\langle\overline{q}q\rangle_{\mbox{\tiny{\rm 0}}}^2{1\over q^2}\ ,
\label{p-1}
\\*[7.2pt]
& &\Pi^q_1(q^2)={4c_1-c_3\over
 32\pi^2}\langle\overline{q}q\rangle_{\mbox{\tiny{\rm 0}}}
\ln(-q^2)
+{c_4+c_5-6c_2\over 96 \pi^2}
\langle g_s\overline{q}\sigma\cdot {\cal G} q
\rangle_{\mbox{\tiny{\rm 0}}}{1\over q^2}
\nonumber
\\*[7.2pt]
& &\hspace*{1.8cm}
+{c_1\over
3}\chi\langle\overline{q}q\rangle_{\mbox{\tiny{\rm 0}}}^2
{1\over q^2}
+{c_1+2c_2\over
24}\chi\langle\overline{q}q\rangle_{\mbox{\tiny{\rm 0}}}
\langle g_s\overline{q}\sigma\cdot {\cal G} q
\rangle_{\mbox{\tiny{\rm 0}}}{1\over (q^2)^2}
\nonumber
\\*[7.2pt]
& &\hspace*{1.8cm}
+ {c_1-2c_2\over
24}\chi_{\mbox{\tiny m}}\langle\overline{q}q\rangle_{\mbox{\tiny{\rm 0}}}
\langle g_s\overline{q}\sigma\cdot {\cal G} q
\rangle_{\mbox{\tiny{\rm 0}}}{1\over (q^2)^2}\; .
\label{p-q}
%
%
%
\\*[14.4pt]
&\Sigma^+ :\hspace{0.7cm}&\Pi^1_1(q^2)=
{3c_2\over 64\pi^2} (q^2)^2\ln(-q^2)
+{3c_2\over 8\pi^2}\chi\langle\overline{q}q
\rangle_{\mbox{\tiny{\rm 0}}}q^2\ln(-q^2)
-{c_1\over 16\pi^2}\chi_{\rm s}
\langle\overline{s}s
\rangle_{\mbox{\tiny{\rm 0}}}q^2\ln(-q^2)
\nonumber
\\*[7.2pt]
& &\hspace*{1.8cm}
-{c_1-2c_3\over 8\pi^2}m_s\langle\overline{q}q
\rangle_{\mbox{\tiny{\rm 0}}}\ln (-q^2)
+{3c_2\over 8\pi^2}m_s
\langle\overline{s}s
\rangle_{\mbox{\tiny{\rm 0}}}\ln (-q^2)
\nonumber
\\*[7.2pt]
& &\hspace*{1.8cm}
-{3c_2\over 16\pi^2}\chi_{\rm m}
\langle g_s\overline{q}\sigma\cdot
{\cal G} q\rangle_{\mbox{\tiny{\rm 0}}}\ln (-q^2)
+{c_2\over 32\pi^2}m_s
\langle g_s\overline{s}\sigma\cdot
{\cal G} s\rangle_{\mbox{\tiny{\rm 0}}}{1\over q^2}
\nonumber
\\*[7.2pt]
& &\hspace*{1.8cm}
+{2c_1+3c_2-6c_3\over 96\pi^2}m_s
\langle g_s\overline{q}\sigma\cdot
{\cal G} q\rangle_{\mbox{\tiny{\rm 0}}}{1\over q^2}
+{c_1-2c_3\over 6}
\langle\overline{q}q
\rangle_{\mbox{\tiny{\rm 0}}}
\langle\overline{s}s
\rangle_{\mbox{\tiny{\rm 0}}}{1\over q^2}
\nonumber
\\*[7.2pt]
& &\hspace*{1.8cm}
+{c_3\over 3}\chi m_s
\langle\overline{q}q
\rangle_{\mbox{\tiny{\rm 0}}}^2{1\over q^2}
+{c_2\over 2}\chi m_s
\langle\overline{q}q
\rangle_{\mbox{\tiny{\rm 0}}}
\langle\overline{s}s
\rangle_{\mbox{\tiny{\rm 0}}}{1\over q^2}
+{c_2\over 2}\chi_{\rm s}m_s
\langle\overline{q}q
\rangle_{\mbox{\tiny{\rm 0}}}
\langle\overline{s}s
\rangle_{\mbox{\tiny{\rm 0}}}{1\over q^2}\ ,
\label{s-1}
\end{eqnarray}
\begin{eqnarray}
& &\Pi^q_1(q^2)=
{3c_2\over 32\pi^4}m_s q^2\ln(-q^2)
+{2c_1-c_3\over 16\pi^2}\langle\overline{q}q
\rangle_{\mbox{\tiny{\rm 0}}}\ln(-q^2)
-{3c_2\over 8\pi^2}
\langle\overline{s}s
\rangle_{\mbox{\tiny{\rm 0}}}\ln(-q^2)
\nonumber
\\*[7.2pt]
& &\hspace*{1.8cm}
+{3c_2\over 8\pi^2}\chi m_s
\langle\overline{q}q
\rangle_{\mbox{\tiny{\rm 0}}}\ln(-q^2)
+{c_3\over 32\pi^2}\chi_{\rm s}m_s
\langle\overline{s}s
\rangle_{\mbox{\tiny{\rm 0}}}\ln(-q^2)
\nonumber
\\*[7.2pt]
& &\hspace*{1.8cm}
+{c_4\over 96\pi^2}
\langle g_s\overline{q}\sigma\cdot
{\cal G} q\rangle_{\mbox{\tiny{\rm 0}}}{1\over q^2}
+{5c_2\over 32\pi^2}
\langle g_s\overline{s}\sigma\cdot
{\cal G} s\rangle_{\mbox{\tiny{\rm 0}}}{1\over q^2}
\nonumber
\\*[7.2pt]
& &\hspace*{1.8cm}
-{7c_2\over 32\pi^2}\chi_{\rm m}m_s
\langle g_s\overline{q}\sigma\cdot
{\cal G} q\rangle_{\mbox{\tiny{\rm 0}}}{1\over q^2}
+{c_5\over 96\pi^2}\chi_{\rm ms} m_s
\langle g_s\overline{s}\sigma\cdot
{\cal G} s\rangle_{\mbox{\tiny{\rm 0}}}{1\over q^2}
\nonumber
\\*[7.2pt]
& &\hspace*{1.8cm}
+{c_1\over 3}\chi
\langle\overline{q}q
\rangle_{\mbox{\tiny{\rm 0}}}^2{1\over q^2}
-c_2\chi
\langle\overline{q}q
\rangle_{\mbox{\tiny{\rm 0}}}
\langle\overline{s}s
\rangle_{\mbox{\tiny{\rm 0}}}{1\over q^2}
-c_2\chi_{\rm s}
\langle\overline{q}q
\rangle_{\mbox{\tiny{\rm 0}}}
\langle\overline{s}s
\rangle_{\mbox{\tiny{\rm 0}}}{1\over q^2}
\nonumber
\\*[7.2pt]
& &\hspace*{1.8cm}
+{c_2\over 2}m_s
\langle\overline{q}q
\rangle_{\mbox{\tiny{\rm 0}}}^2{1\over (q^2)^2}
+{c_3-2c_1\over 12}m_s
\langle\overline{q}q
\rangle_{\mbox{\tiny{\rm 0}}}
\langle\overline{s}s
\rangle_{\mbox{\tiny{\rm 0}}}{1\over (q^2)^2}
\nonumber
\\*[7.2pt]
& &\hspace*{1.8cm}
+{c_1\over 24}\chi
\langle\overline{q}q
\rangle_{\mbox{\tiny{\rm 0}}}
\langle g_s\overline{q}\sigma\cdot
{\cal G} q\rangle_{\mbox{\tiny{\rm 0}}}
{1\over (q^2)^2}
-{5c_2\over 24}\chi
\langle\overline{q}q
\rangle_{\mbox{\tiny{\rm 0}}}
\langle g_s\overline{s}\sigma\cdot
{\cal G} s\rangle_{\mbox{\tiny{\rm 0}}}
{1\over (q^2)^2}
\nonumber
\\*[7.2pt]
& &\hspace*{1.8cm}
-{7c_2\over 24}\chi_{\rm s}
\langle\overline{s}s
\rangle_{\mbox{\tiny{\rm 0}}}
\langle g_s\overline{q}\sigma\cdot
{\cal G} q\rangle_{\mbox{\tiny{\rm 0}}}
{1\over (q^2)^2}
+{c_1\over 24}\chi_{\rm m}
\langle\overline{q}q
\rangle_{\mbox{\tiny{\rm 0}}}
\langle g_s\overline{q}\sigma\cdot
{\cal G} q\rangle_{\mbox{\tiny{\rm 0}}}
{1\over (q^2)^2}
\nonumber
\\*[7.2pt]
& &\hspace*{1.8cm}
-{5c_2\over 24}\chi_{\rm ms}
\langle\overline{q}q
\rangle_{\mbox{\tiny{\rm 0}}}
\langle g_s\overline{s}\sigma\cdot
{\cal G} s\rangle_{\mbox{\tiny{\rm 0}}}
{1\over (q^2)^2}
-{7c_2\over 24}\chi_{\rm m}
\langle\overline{s}s
\rangle_{\mbox{\tiny{\rm 0}}}
\langle g_s\overline{q}\sigma\cdot
{\cal G} q\rangle_{\mbox{\tiny{\rm 0}}}
{1\over (q^2)^2}\ .
\label{s-q}
%
\\*[14.4pt]
&\Xi^0 :\hspace{0.7cm}&\Pi^1_1(q^2)=-{c_1\over 128
 \pi^4}(q^2)^2\ln(-q^2)
-{c_1\over
16\pi^2}\chi\langle\overline{q}q\rangle_{\mbox{\tiny{\rm 0}}}
q^2\ln(-q^2)
+{3c_2\over 8\pi^2}\chi_{\rm s}
\langle\overline{s}s\rangle_{\mbox{\tiny{\rm 0}}}
q^2\ln(-q^2)
\nonumber
\\*[7.2pt]
& &\hspace*{1.8cm}
+{3c_2\over 8\pi^2}m_s
\langle\overline{q}q\rangle_{\mbox{\tiny{\rm 0}}}
\ln(-q^2)
-{c_1-2c_3\over 8\pi^2}m_s
\langle\overline{s}s\rangle_{\mbox{\tiny{\rm 0}}}
\ln(-q^2)
\nonumber
\\*[7.2pt]
& &\hspace*{1.8cm}
-{3c_2\over 16\pi^2}\chi_{\rm ms}
\langle g_s\overline{s}\sigma\cdot
{\cal G} s\rangle_{\mbox{\tiny{\rm 0}}}
\ln(-q^2)
+{c_2\over 32\pi^2}m_s
\langle g_s\overline{q}\sigma\cdot
{\cal G} q\rangle_{\mbox{\tiny{\rm 0}}}{1\over q^2}
\nonumber
\\*[7.2pt]
& &\hspace*{1.8cm}
+{2c_1+3c_2-6c_3\over 96\pi^2}m_s
\langle g_s\overline{s}\sigma\cdot
{\cal G} s\rangle_{\mbox{\tiny{\rm 0}}}{1\over q^2}
-{c_3\over 6}
\langle\overline{s}s\rangle_{\mbox{\tiny{\rm 0}}}^2
{1\over q^2}
-{c_2\over 2}
\langle\overline{s}s\rangle_{\mbox{\tiny{\rm 0}}}
\langle\overline{q}q\rangle_{\mbox{\tiny{\rm 0}}}
{1\over q^2}
\nonumber
\\*[7.2pt]
& &\hspace*{1.8cm}
-{c_1-2c_3\over 6}\chi m_s
\langle\overline{s}s\rangle_{\mbox{\tiny{\rm 0}}}
\langle\overline{q}q\rangle_{\mbox{\tiny{\rm 0}}}
{1\over q^2}
-{c_1-2c_3\over 6}\chi_{\rm s}m_s
\langle\overline{s}s\rangle_{\mbox{\tiny{\rm 0}}}
\langle\overline{q}q\rangle_{\mbox{\tiny{\rm 0}}}
{1\over q^2}\ ,
\label{x-1}
\end{eqnarray}
\begin{eqnarray}
& &\Pi^q_1(q^2)=
{3c_2\over 32\pi^4}m_s q^2\ln(-q^2)
-{c_3\over
 32\pi^2}\langle\overline{q}q\rangle_{\mbox{\tiny{\rm 0}}}
\ln(-q^2)
-{3c_2\over 8\pi^2}
\langle\overline{s}s\rangle_{\mbox{\tiny{\rm 0}}}
\ln(-q^2)
\nonumber
\\*[7.2pt]
& &\hspace*{1.8cm}
+{3c_2\over 8\pi^2}\chi m_s
\langle\overline{q}q\rangle_{\mbox{\tiny{\rm 0}}}
\ln(-q^2)
-{2c_1-c_3\over 16\pi^2}\chi_{\rm s}m_s
\langle\overline{s}s\rangle_{\mbox{\tiny{\rm 0}}}
\ln(-q^2)
\nonumber
\\*[7.2pt]
& &\hspace*{1.8cm}
-{c_5\over 96\pi^2}\langle g_s\overline{q}\sigma\cdot
{\cal G} q\rangle_{\mbox{\tiny{\rm 0}}}{1\over q^2}
+{7c_2\over 32\pi^2}
\langle g_s\overline{s}\sigma\cdot
{\cal G} s\rangle_{\mbox{\tiny{\rm 0}}}{1\over q^2}
\nonumber
\\*[7.2pt]
& &\hspace*{1.8cm}
-{5c_2\over 32\pi^2}\chi_{\rm m}m_s
\langle g_s\overline{q}\sigma\cdot
{\cal G} q\rangle_{\mbox{\tiny{\rm 0}}}
{1\over q^2}
-{c_4\over 96\pi^2}\chi_{\rm ms}m_s
\langle g_s\overline{s}\sigma\cdot
{\cal G} s\rangle_{\mbox{\tiny{\rm 0}}}
{1\over q^2}
\nonumber
\\*[7.2pt]
& &\hspace*{1.8cm}
-c_2\chi
\langle\overline{q}q\rangle_{\mbox{\tiny{\rm 0}}}
\langle\overline{s}s\rangle_{\mbox{\tiny{\rm 0}}}
{1\over q^2}
-c_2\chi_{\rm s}
\langle\overline{q}q\rangle_{\mbox{\tiny{\rm 0}}}
\langle\overline{s}s\rangle_{\mbox{\tiny{\rm 0}}}
{1\over q^2}
+{c_1\over 3}\chi_{\rm s}
\langle\overline{s}s\rangle_{\mbox{\tiny{\rm 0}}}^2
{1\over q^2}
\nonumber
\\*[7.2pt]
& &\hspace*{1.8cm}
+{c_2\over 2}m_s
\langle\overline{s}s\rangle_{\mbox{\tiny{\rm 0}}}^2
{1\over (q^2)^2}
+{c_3-2c_1\over 12}m_s
\langle\overline{q}q\rangle_{\mbox{\tiny{\rm 0}}}
\langle\overline{s}s\rangle_{\mbox{\tiny{\rm 0}}}
{1\over (q^2)^2}
\nonumber
\\*[7.2pt]
& &\hspace*{1.8cm}
-{7c_2\over 24}\chi
\langle\overline{q}q\rangle_{\mbox{\tiny{\rm 0}}}
\langle g_s\overline{s}\sigma\cdot
{\cal G} s\rangle_{\mbox{\tiny{\rm 0}}}{1\over (q^2)^2}
+{c_1\over 24}\chi_{\rm s}
\langle\overline{s}s\rangle_{\mbox{\tiny{\rm 0}}}
\langle g_s\overline{s}\sigma\cdot
{\cal G} s\rangle_{\mbox{\tiny{\rm 0}}}{1\over (q^2)^2}
\nonumber
\\*[7.2pt]
& &\hspace*{1.8cm}
-{5c_2\over 24}\chi_{\rm s}
\langle\overline{s}s\rangle_{\mbox{\tiny{\rm 0}}}
\langle g_s\overline{q}\sigma\cdot
{\cal G} q\rangle_{\mbox{\tiny{\rm 0}}}{1\over (q^2)^2}
-{5c_2\over 24}\chi_{\rm m}
\langle\overline{s}s\rangle_{\mbox{\tiny{\rm 0}}}
\langle g_s\overline{q}\sigma\cdot
{\cal G} q\rangle_{\mbox{\tiny{\rm 0}}}{1\over (q^2)^2}
\nonumber
\\*[7.2pt]
& &\hspace*{1.8cm}
+{c_1\over 24}\chi_{\rm ms}
\langle\overline{s}s\rangle_{\mbox{\tiny{\rm 0}}}
\langle g_s\overline{s}\sigma\cdot
{\cal G} s\rangle_{\mbox{\tiny{\rm 0}}}{1\over (q^2)^2}
-{7c_2\over 24}\chi_{\rm ms}
\langle\overline{q}q\rangle_{\mbox{\tiny{\rm 0}}}
\langle g_s\overline{s}\sigma\cdot
{\cal G} s\rangle_{\mbox{\tiny{\rm 0}}}{1\over (q^2)^2}
\; .
\label{x-q}
\end{eqnarray}
Here $c_1$, $c_2$, $c_3$, $c_4$, and $c_5$ have been
defined in Eq.~(\ref{c-def}), and we have ignored the isospin
 breaking in the vacuum condensates (i.e.,
$\langle\overline{u}\hat{O}u\rangle_{\mbox{\tiny{\rm 0}}}
\simeq\langle\overline{d}\hat{O}d\rangle_{\mbox{\tiny{\rm 0}}}
=\langle\overline{q}\hat{O}q\rangle_{\mbox{\tiny{\rm 0}}}$).
All polynomials in $q^2$,
which vanish under the Borel transformation, have been omitted
in Eqs.~(\ref{p-1}--\ref{x-q}).
\newpage

%
%
\begin{figure}
\caption{Optimized sum-rule prediction for $H_p$ as function of
$\chi$, with Ioffe's interpolating field (i.e., $t=-1$).
The other input parameters are described in the text.}
\label{fig-1}
\end{figure}
\begin{figure}
\caption{The left-hand side (solid) and right-hand side
(long-dashed) of Eq.~(\protect{\ref{sum_p_q}}) as functions
of Borel $M^2$ for $t=-1$, with $\chi=1.8\,\text{GeV}^{-1}$
and the optimized
values for $H_p$, $\Delta\lambda_p^2$, and $\Delta s_0^q$.
The curves 1, 2, and 3 correspond to the first, second,
and third terms on the right-hand side of
Eq.~(\protect{\ref{sum_p_q}}).}
\label{fig-2}
\end{figure}
\begin{figure}
\caption{Optimized sum-rule prediction for $H_{\Sigma^+}$
as function of $\chi$, with $t=-1$. The three curves
correspond to $\chi_{\rm s}=0$ (solid),
$1.5\,\text{GeV}^{-1}$ (dashed), and $3.0\,\text{GeV}^{-1}$
(dotted). The other input parameters are the same as in
Fig.~\protect{\ref{fig-1}}.}
\label{fig-3}
\end{figure}
\begin{figure}
\caption{Optimized sum-rule prediction for $H_{\Xi^0}$
as function of $\chi_{\rm s}$, with $t=-1$.
The other input parameters are the same as in
Fig.~\protect{\ref{fig-1}}.}
\label{fig-4}
\end{figure}
\begin{figure}
\caption{Optimized sum-rule prediction for $H_p$,
$H_{\Sigma^+}$, and $H_{\Xi^0}$
as functions of $t$, with $\chi=2.5\,\text{GeV}^{-1}$
and $\chi_{\rm s}=1.5\,\text{GeV}^{-1}$.
The other input parameters are the same as in
Fig.~\protect{\ref{fig-1}}.}
\label{fig-5}
\end{figure}
\begin{figure}
\caption{Optimized sum-rule prediction for $H_p$
and $H_{\Sigma^+}$ as functions of $\chi$,
with $t=-1$ and $\chi_{\rm s}=\chi_{\rm ms}
=1.5\,\text{GeV}^{-1}$. The three curves
correspond to $\chi_{\rm m}=\chi$ (solid),
${1\over 2}\chi$ (dashed),
and ${3\over 2}\chi$ (dotted).
The other input parameters are the same as in
Fig.~\protect{\ref{fig-1}}.}
\label{fig-6}
\end{figure}
\begin{figure}
\caption{Optimized sum-rule prediction for $H_{\Sigma^+}$
and $H_{\Xi^0}$ as functions of $\chi_{\rm s}$,
with $t=-1$ and $\chi=\chi_{\rm m}
=2.5\,\text{GeV}^{-1}$. The three curves
correspond to $\chi_{\rm ms}=\chi_{\rm s}$ (solid),
${1\over 2}\chi_{\rm s}$ (dashed),
and ${3\over 2}\chi_{\rm s}$ (dotted).
The other input parameters are the same as in
Fig.~\protect{\ref{fig-1}}.}
\label{fig-7}
\end{figure}
\begin{figure}
\caption{Optimized sum-rule prediction for $H_p$
as functions of $\chi$, with $t=-1$. The solid
curve is obtained by including all three terms
on the RHS of Eq.~(\protect{\ref{sum_p_q}}),
while the dashed curve is obtained by neglecting
the third term on the RHS of Eq.~(\protect{\ref{sum_p_q}}).}
\label{fig-8}
\end{figure}

\end{document}